\documentclass[aps,prl,twocolumn,showpacs,showkeys,groupedaddress,floatfix,amsmath,amssymb]{revtex4}
\usepackage{graphicx}
\usepackage{dcolumn}
\usepackage{bm}

\begin{document}


\title{Current-voltage characteristics of quasi-one-dimensional
superconductors: An S-curve in the constant voltage regime}

\author{D.Y. Vodolazov$^1$}
\author{F.M. Peeters$^1$}
\email{peeters@uia.ua.ac.be}
\author{L. Piraux$^2$}
\author{S. M\'at\'efi-Tempfli$^2$}
\author{S. Michotte$^2$}
\affiliation{
$^1$ Departement Natuurkunde, Universiteit Antwerpen
(UIA), B-2610 Antwerpen,Belgium\\
$^2$Unit\'e de Physico-Chimie et de Physique des Mat\'eriaux
(PCPM), Universit\'e catholique de Louvain (UCL), Place Croix du
Sud 1, B-1348 Louvain-la-Neuve, Belgium}

\date{\today}

\begin{abstract}
Applying a constant voltage to superconducting nanowires we find
that its IV-characteristic exhibits an unusual S-behavior. This
behavior is the direct consequence of the dynamics of the
superconducting condensate and of the existence of two different
critical currents: $j_{c2}$ at which the pure superconducting
state becomes unstable and $j_{c1}< j_{c2}$ at which the phase
slip state is realized in the system.
\end{abstract}

\pacs{74.25.Op, 74.20.De, 73.23.-b}

\maketitle

The majority of the experiments on the resistive state in
quasi-one dimensional systems were performed in the constant
current regime and at temperatures close to $T_c$. It is extremely
difficult to apply voltage to a superconductor because the current
density induced by the applied electric field inevitably reaches
the critical value and destroys superconductivity in the sample.
The decrease of the superconducting current by the appearance of
phase slip centers
\cite{Skocpol,Kramer1,Kramer2,Kramer3,Watts-Tobin} is not
effective in this case because of the large heating of the sample
at low temperatures. At temperatures close to $T_c$ the heating
can be suppressed due to the low value of the critical currents
but in this case the applied electric field does not penetrate
deep into the sample because of the existence of regions near the
N-S boundary where the drop of the applied voltage occurs
\cite{Yu,Ryazanov}.

This situation drastically changes with the appearance of
nano-technology and the ability to create long (to allow the
appearance of phase slip centers) superconducting wires with a
small cross section (to decrease the effect of heating). In this
Letter we present results on the behavior of such nanowires in the
constant voltage regime. We found that the I-V characterestic in
this case has a remarkable S-shape. Our theoretical analysis based
on the time-dependent Ginzburg-Landau equations (TDGL) shows that
such a behavior is a direct consequence of the dynamics of the
superconducting condensate and we predict new unusual features
which still need additional experimental study.

The superconducting nanowires were prepared by electrodeposition
into nanopores of homemade track-etched polycarbonate
membranes\cite{Legras}. For the lead nanowires, a 22 $\mu$m thick
membrane (with pore diameter $\sim$ 40 nm and pore density $\sim$
4$\cdot$10$^{9}$ cm$^{-2}$) and an aqueous solution of 40.4 g/l
Pb(BF$_{4}$)$_{2}$, 33.6 g/l HBF$_{4}$ and 15 g/l H$_{3}$BO$_{3}$
were used \cite{Michotte}, while in the case of the tin nanowires,
a 50 $\mu$m thick membrane (with pore diameter $\sim$ 55 nm and
pore density $\sim$ 2$\cdot$10$^{9}$ cm$^{-2}$) and an electrolyte
of 41.8 g/l Sn(BF$_{4}$)$_{2}$ in water solution were applied.
Constant potential of -0.5 V versus an Ag/AgCl reference electrode
was used in a three-electrode configuration in order to reduce the
Pb$^{2+}$ or Sn$^{2+}$ ions into the nanopores. As shown in Fig.
1, the nanowires are cylindrical and the diameter is uniform along
their length.
\begin{figure}[hbtp]
\includegraphics[width=0.38\textwidth]{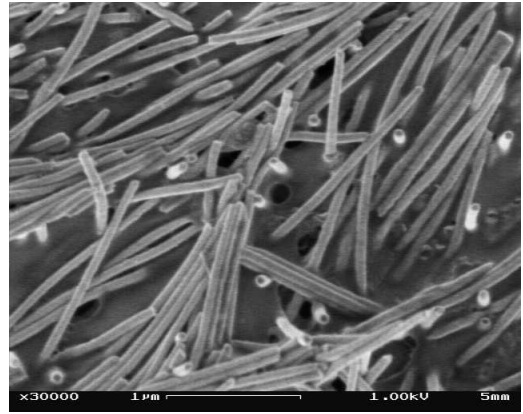}
\caption{SEM-FEG micrograph of lead nanowires after dissolution of
the hosting polycarbonate membrane with dichloromethane.}
\end{figure}
In order to perform electrical transport measurements on a single
nanowire (still kept inside the membrane), a self-contacting
technique was developed. In addition of the thick gold cathode
($\sim$ 1 $\mu$m) from which the nanowires start to grow up inside
the membrane, another thin gold layer (in the range of 50 to 200
nm) was deposited on the other side exposed to the electrolyte
prior to the electrodeposition. This contacting layer covers the
surface but in contrast to the cathode, it is not thick enough to
plug the pores. As the nanowires do not grow at the same speed,
the first emerging nanowire interrupts the growth of the others by
favoring the formation of a film of the deposited material on this
contacting layer.

All the transport measurements are done in an electromagnetically
shielded environment and special precautions were taken to reduce
the noise. The voltage across the sample was measured by a
Keithley voltmeter. From the numerous measurements performed both
on Pb and on Sn nanowires, it appears that almost no depression of
T$_{c}$ was observed. In addition, a superconducting transition
was found to occur when the residual resistivity ratio (RRR) is
larger than about 2 while for smaller RRR values an insulating
behavior is observed \cite{Michotte}. Our nanowires are moderately
disordered and the RRR of the samples was larger than 10.
Structural characterization of the Pb nanowires using TEM
experiments confirm also their polycrystalline character
\cite{Dubois}.

In Fig. 2, the DC current-voltage (I-V) characteristics of the Pb
and Sn nanowires are reported, applying either a DC current or a
DC voltage.
\begin{figure}[t]
\includegraphics[width=0.41\textwidth]{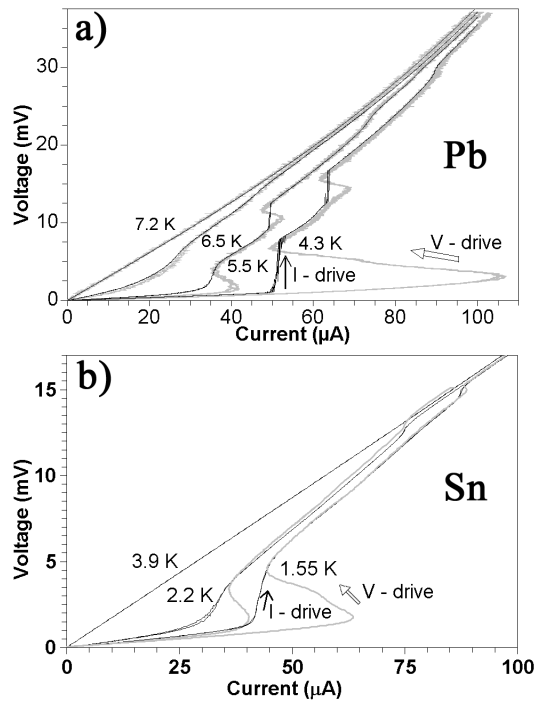}
\caption{Current-voltage characteristics at different temperatures
of a) a Pb nanowire (diameter 40 nm, length 22 $\mu$m) and b) a Sn
nanowire (diameter 55 nm, length 50 $\mu$m). Results are shown for
the current driven mode (black curves) and the voltage driven mode
(grey curves).}
\end{figure}
All the curves presented in Fig. 2 were measured in both
directions and no hysteresis was observed. These measurements were
also performed with reversed polarity without any changes compared
to the reported results. In the usual current driven experiment we
observe for the Pb nanowire (Fig. 2(a)) the successive appearance
of two PSCs at low temperature. Applying the phenomenological SBT
model \cite{Skocpol}, the size of the first PSC can be estimated
to be about 18 $\mu$m, which is twice the quasiparticles diffusion
length. In spite that successive PCSs tend to avoid those already
in place, the second PSC created in this Pb nanowire is thus
forced to interpenetrate the first one, which explains why the
second jump in resistance is smaller. The current driven
experiment on the 50 $\mu$m long Sn nanowire (Fig. 2(b)) also
shows the formation of a PSC at low temperature but for this
particular material, the PSC extension is larger (around 40
$\mu$m) and the formation of a second PSC almost coincides with
the transition to the normal state. Interesting new features were
observed when measuring the reverse way, i.e. applying constant
voltage and measuring the current. Here, the current flowing into
the nanowire was determined by measuring the voltage across a 1
$\Omega$ resistance added in series and the voltage developed
across the sample was measured separately. In this voltage driven
experiment, we found an interesting S-shape behavior which occurs
at low temperature in the formation region of these PSCs, both for
the Pb and the Sn nanowires.

To understand the experimental results we used the generalized
TDGL equation \cite{Kramer1} which describes the dynamics of the
superconducting condensate $\psi=|\psi|e^{i\theta}$
\begin{equation}
\frac{u}{\sqrt{1+\gamma^2|\psi|^2}} \left(\frac
{\partial}{\partial t}+i\varphi+\frac{\gamma^2}{2}\frac{\partial
|\psi|^2}{\partial t} \right)\psi=\frac{\partial^2 \psi}{\partial
s^2}+\psi(1-|\psi|^2).
\end{equation}
The physical quantities are presented in dimensionless units (see
e.g. Ref. \cite{Kramer2}) and we neglected the vector potential,
because  in our case the self-induced magnetic field is negligible
in size. Eq. (1) is supplemented with an equation for the
electrostatic potential $\varphi$ which is obtained from the
condition for the conservation of the total current in the wire,
i.e. ${\rm div} {\bf j}=0$. The parameter $\gamma$ depends on the
material and we took $u\simeq 5.79$ \cite{Kramer1}.

Although Eq. (1) is valid only close to $T_c$ (see \cite{Kramer1})
we expect to find clues of the dynamics of the system even far
from $T_c$. This expectation is motivated by the fact that results
found from the solution of the stationary Ginzburg-Landau
equations for mesoscopic superconductors \cite{Deo} agree with
experiment even far from $T_c$.
\begin{figure}[t]
\includegraphics[width=0.48\textwidth]{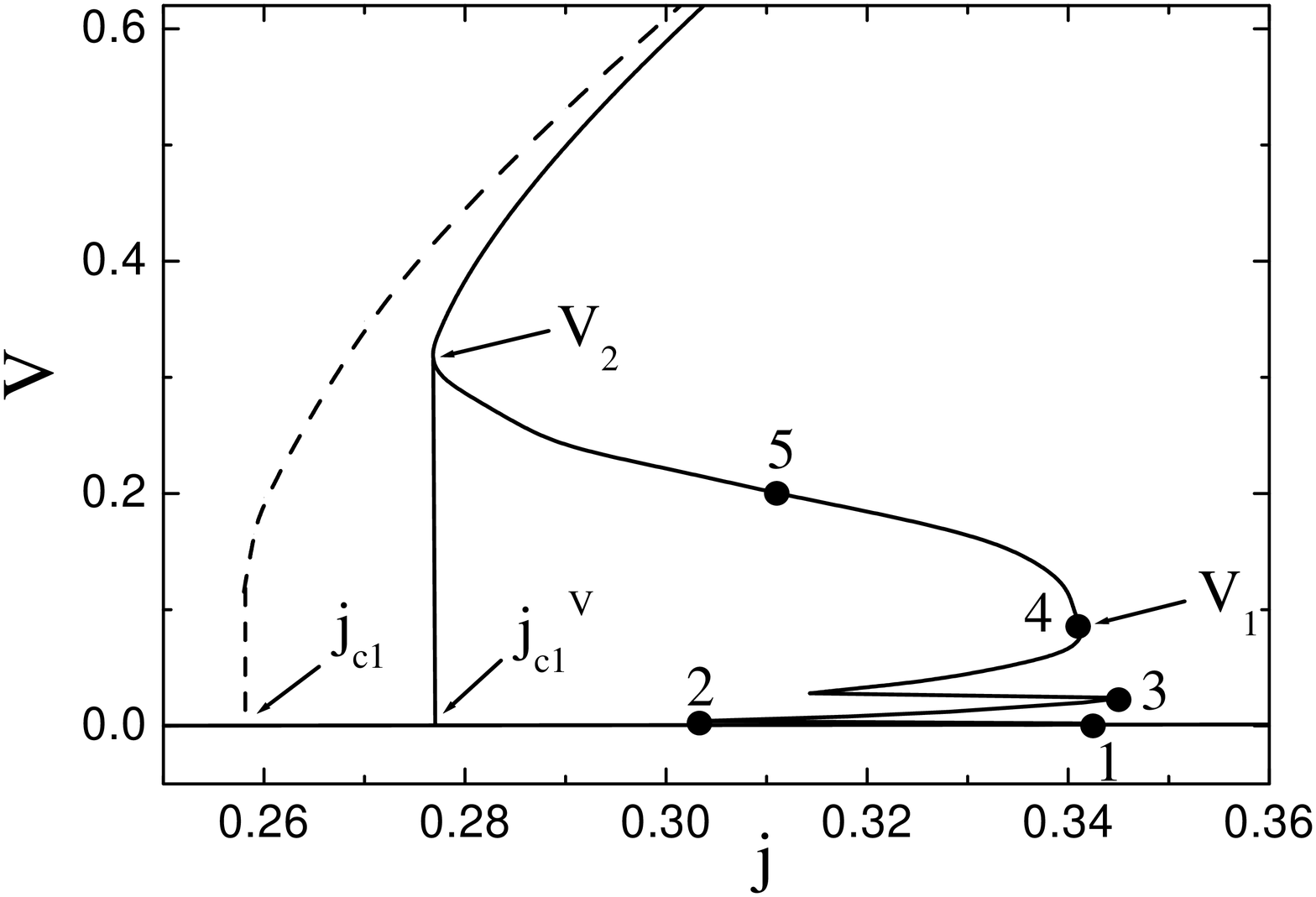}
\caption{The current voltage characteristics of a superconducting
wire of length $L=40\xi$ and $\gamma=10$ ($\lambda_Q\simeq
2.3\xi$). Dotted(solid) curve for the $I(V)=const$ regime.}
\end{figure}

If we apply an external current to the wire then the system
exhibits hysteretic behavior \cite{Kramer1,Kramer3}. If one starts
from the superconducting state and increases the current the
superconducting state switches to the resistive superconducting or
normal state at the upper critical current $j_{c2}$. With
decreasing current it is possible to keep the sample in this state
even for currents up to $j_{c1}<j_{c2}$ (which we call the lower
critical current). The superconducting resistive state is realized
as a periodic oscillation of the order parameter in time at one
point of the superconductor \cite{Kramer1,Kramer2}. When the order
parameter reaches zero in this point, a phase slip of $2\pi$
occurs. Such a state is now called a phase slip state and the
corresponding point a phase slip center (PSC). The meaning of the
current $j_{c1}$ is that for $j<j_{c1}$ the phase slip solution
cannot be realized (in the absence of fluctuations) and thus the
current $j_{c1}$ is the critical current at which PSC's start to
appear. Thus there is a limiting cycle (see Ref. \cite{Ivlev}) for
Eq. (1) giving rise to this process.

The existence of such two critical currents is, in the $V=const$
regime, responsible for a complicated I-V characteristic (see Fig.
3). The I(V) curve has an overall S-like behavior which at low
voltages exhibits an oscillatory behavior superimposed on it. We
found that these unusual properties are typical for
superconducting wires with $\gamma \gg 1$. To understand the
physics leading to this unusual behavior we consider the dynamics
of $\psi$ in the wire for different voltages.

When we apply a voltage $V$ to the wire of length $L$ a constant
electrical field appears in the sample $E=j_E=V/L$. As a result
the superconducting condensate will be accelerated and the
momentum $p=\nabla \theta$ will increase in time. For small
voltages and electric fields the absolute value of the order
parameter and current density are approximately described by
quasi-equilibrium expressions with $|\psi|(t)=1-p(t)^2$ and
$j_s(t)=p(t)(1-p(t)^2)$ ($p(t)\simeq j_Et$) while $p<p_c$ (at
$L/\xi \gg 1$ we have $p_c\simeq 1/\sqrt{3}$). For $p>p_c$ the
above spatial independent solution becomes 'unstable' (like for
the case of a ring in the presence of an applied magnetic field as
discussed in Refs. \cite{Tarlie,Vodolazov}). The order parameter
vanishes in the center of the wire and a jump of the phase by
$2\pi$ occurs in this point. It decreases $p$ by $\Delta p=2\pi/L$
(and hence the superconducting current density) in the entire
wire. A second phase slip will appear in the same point and the
phase slip process repeats itself as long as $j>j_{c1}$. When the
current density becomes less than $j_{c1}$ the order parameter
will increase till its quasi-equilibrium value
$|\psi|(t)=1-p(t)^2$ with $p(t)<p_c$. Than the applied electric
field will again accelerate the superconducting condensate and the
above process is repeated (see Fig. 4). We define the time which
is necessary to decrease the order parameter from
$|\psi|=\sqrt{2/3}$ till zero and again back to the value
$|\psi|=\sqrt{2/3}$ as the transition time $T_{tr}$ and call this
entire process the {\it transition process}.
\begin{figure}[t]
\includegraphics[width=0.48\textwidth]{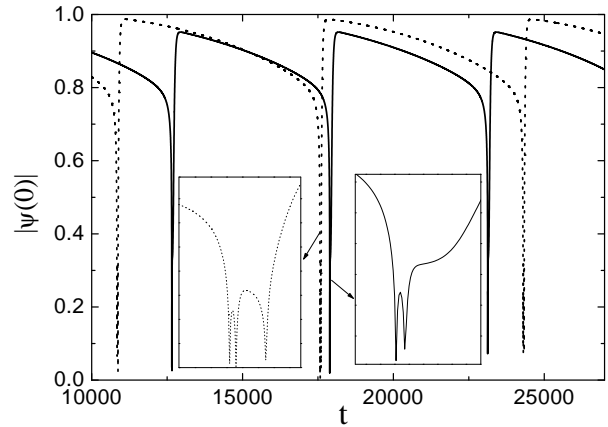}
\caption{Dependence of the order parameter in the center of the
wire (in the minimal point of $|\psi(s)|$) as function of time.
Solid(dotted) curve corresponds to the voltage in point 1(2)-
$V=0.0024$ ($V=0.0028$) of Fig. 3.}
\end{figure}

It is easy to show that at small voltages (in the limit $V\to 0$)
the number of phase slips which occurs during the transition
process in an ideal wire is given by the following expression
${\rm N_{min}=Nint\left((p_c-p_{c1})(L/2\pi+1)\right)}$, where
$p_{c1}$ is the smallest real root of equation
$j_{c1}=p_{c1}(1-p_{c1}^2)$ and $\rm{Nint(x)}$ returns the nearest
integer value. The number of phase slips occurring during the
transition process increases with increasing wire length.

Because of the presence of an external electric field, $p$ also
increases during the transition process. Roughly speaking, if the
voltage in the wire is so large that $j_ET_{tr}>2\pi /L$ the
number of phase slips during the transition will increase by one
(see Fig. 4). It leads to an instant decrease of the average
current. The reason is that the maximum current during current
oscillations $j_{max}$ for a small change of $V$ does not change
but the minimum value of the current $j_{min}$ decreases by about
$2\pi/L$, and hence $<I>$ should decrease. This effect is most
pronounced in short wires where $N_{min}$ is small and the term
$2\pi/L$ is large. If $N$ does not change with increasing $V$ then
$<I>$ increases because both minimum and maximum currents grow.

When the period $T=2\pi N/V$ becomes of order $2T_{tr}$ the
dynamics of the system changes considerably (at voltage $V_1$ in
Fig. 3). Starting from this point the average current $<I>$
monotonically decreases with increasing applied voltage. For these
values of the voltages the system is unable to return to the
"quasi-equilibrium" state (it is a state such that if we switch
off the voltage instantly the system remains in this state for an
infinitely long time in the absence of fluctuations). At any
moment of time the order parameter is more suppressed in the point
where the oscillations occur. At these voltages $j_{max}$
decreases with increasing $V$ and in principle it should reach the
current $j_{c1}$ (and $j_{min}$ should reach $j_{c1}$ from the
bottom). But this is only possible in an infinitely long wire. In
a finite length wire every phase slip event leads to a decrease of
the momentum $p$ and the current $j$ by a finite value. As a
result the lower critical current in the $V=const$ regime is equal
to $j_{c1}^V=j_{c1}+dj$ (with $dj\sim 1/L$) at which the 'usual',
like in the $I=const$ regime, periodic phase slip processes with
period $T=2\pi/<V>$ starts. This corresponds to the voltage $V_2$
in Fig. 3.

In long wires a complex behavior is found in the current
decreasing region due to the existence of several phase slip
events during the transition process. At the initial parts of this
region an irregular behavior is typical. In Fig. 5 we show (dotted
curve) an example of such a situation (at $V=0.08$ - point 4 in
Fig. 3). Not a single, but several periods $T=2\pi N/V$ are
present. This regime is replaced at higher voltages (for example
at $V=0.2$ - point 5 in Fig. 3) by a regime with a single period
and the situation becomes similar to the case at low voltages.
\begin{figure}[t]
\includegraphics[width=0.48\textwidth]{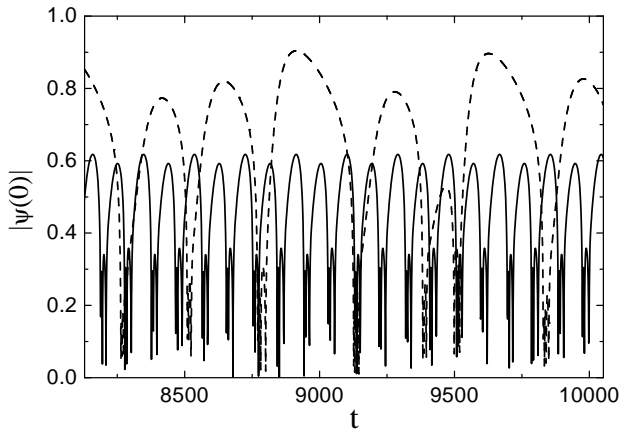}
\caption{Time dependence of the order parameter in the current
decreasing regime for a long wire ($L=40\xi$). Dotted(solid) curve
corresponds to the voltage in point 4(5)- $V=0.08$ ($V=0.2$) of
Fig. 3.}
\end{figure}

We expect that the S shape of the I-V characteristic in the
$V=const$ regime will be conserved even for temperatures far from
$T_c$, because the only condition for its occurrence is the
existence of two different critical currents $j_{c1}<j_{c2}$.
While $V<V_1$ the current in the wire will be an oscillating
function with maximal amplitude $j_{c2}$ and minimal amplitude
$j_{c1}$. It means that the time averaged current should be larger
than $j_{c1}$. At $V>V_1$ we expect that $j_{max}$ decreases due
to the same mechanisms as was discussed above and hence the
average current should decrease. But the maximal current cannot
decrease below $j_{c1}$ and for $V>V_2$ both $j_{min}$ and
$j_{max}$ have to increase with increasing voltage.

In our theoretical consideration we neglected the effect of
thermo- and/or quantum-activated phase-slip centers. We believe
that this is the reason why in the experiment no hysteresis was
observed in the $I=const$ regime. The larger these fluctuations
the larger they will decrease the measured $j_{c2}^{exp}$ and
increase the measured $j_{c1}^{exp}$ (see also the discussion of
this question in Ref. \cite{Kramer4}) and in principle they may
suppress the hysteresis. But in the $V=const$ regime the situation
is quite different. If the time needed for the appearance of a
fluctuating phase slip center is much larger than the period of
oscillation of the current then we can neglect their effect.

In conclusion, we presented the first experiment on the I-V
characteristic in the voltage constant regime for narrow
superconducting Pb and Sn wires. An unexpected S-shape response is
found which is absent in a current driven measurements. A
theoretical analysis based on the generalized TDGL equations gives
a qualitative explanation of this phenomena which is driven by a
highly nonlinear time response of the superconductor to electric
fields leading to the periodic creation of PS centers.
Furthermore, we predict an oscillatory I-V characteristic at low
voltages which was not found in the experiment presumedly due to
the large length of the wire and hence the relatively large value
of $N_{min}$.

This work was supported by IUAP(P5/1/1), GOA, ESF, FWO-Vl and the
"Communaut\'e Fran\c caise de Belgique" through the program
"Actions de Recherches Concert\'ees". We thank the POLY lab. at
UCL for the polymer membranes. S. Mi. is a Research Fellow of the
FNRS.

\end{document}